\documentclass[aps,prb,floats,twocolumn]{revtex4-1}
\usepackage{graphicx}
\usepackage{amssymb}
\usepackage{color}
\graphicspath{{../../FIG_FINALI/}}

\begin{document}
\title{Transport properties of armchair graphene nanoribbon junctions
between graphene electrodes }

\author{C. Motta}
\email[]{carlomotta84@gmail.com}
\affiliation{Dipartimento di Scienza dei Materiali, Universit\`a di Milano-Bicocca, Via Cozzi 53, 20125 Milano, Italy}

\author{D. S\'anchez-Portal}
\affiliation{Centro de F\'{i}sica de Materiales CSIC-UPV/EHU, Paseo
Manuel de Lardizabal 5, 20018 Donostia-San Sebasti\'an, Spain}
\affiliation{Donostia International Physics Center (DIPC), Paseo Manuel de Lardizabal 4, 20018 Donostia-San Sebastian, Spain}

\author{M. I. Trioni}
\affiliation{CNR - National Research Council of Italy, ISTM, Via Golgi 19, 20133 Milano, Italy}
\affiliation{Donostia International Physics Center (DIPC), Paseo Manuel de Lardizabal 4, 20018 Donostia-San Sebastian, Spain}
\date{\today}

\begin{abstract}
The transmission properties of armchair graphene nanoribbon junctions between graphene
electrodes are investigated by means of first-principles quantum transport calculations.
First the dependence of the transmission function on the size of the nanoribbon has been
studied.
Two regimes are highlighted: for small applied bias transport takes place via tunneling 
and the length of the ribbon is the key parameter that determines the junction conductance;
at higher applied bias resonant transport through HOMO and LUMO starts to play a more
determinant role, and the transport properties depend on the details of  the geometry (width
and length) of the carbon nanoribbon.
In the case of the thinnest ribbon it has been verified that a tilted geometry of the central
phenyl ring is the most stable configuration.
As a consequence of this rotation the conductance decreases due to the misalignment of
the $pi$ orbitals between the phenyl ring and the remaining part of the junction.
All the computed transmission functions have shown a negligible dependence on different
saturations and reconstructions of the edges of the graphene leads, suggesting a general
validity of the reported results.
\end{abstract}
\maketitle
\section{Introduction}
\label{sec:intro}
Graphene, a perfect carbon monolayer $sp^2-$hybridized, has attracted a huge interest since
its discovery.\cite{Novoselov_Geim_Morozov_Jiang_Zhang_Dubonos_Grigorieva_Firsov_2004,Geim_Novoselov_2007}
Besides a pure theoretical interest, its possible applications in carbon-based electronics
represent a very exciting perspective.
Graphene displays a very peculiar electronic structure, arising from the confinement of electrons
in two dimensions and its geometrical symmetries.
Graphene is a zero gap semimetal whose specific linear electronic band dispersion near the
Brillouin zone corners (Dirac points) gives rise to electrons and holes that propagate as massless
fermions.\cite{Cast09,Dubois_Zanolli_Declerck_Charlier_2009,PhysRevB.44.13237,Wallace_1947}
Graphene nanoribbons (GNRs) are one dimensional graphene strips that are considered as
promising candidates building blocks for future electronic applications.\cite{Avouris_Chen_2007,%
Wang_Ang_Wang_Tang_Thong_Loh_2010,Wakabayashi_Fujita_Ajiki_Sigrist_1998}
Several methods allow the production of GNR, including mechanical cutting of exfoliated
graphene,\cite{Geim_Novoselov_2007} patterning of epitaxially grown
graphene,\cite{Berger_Song_Li_Wu_Brown_Naud_Mayou_Li_Hass_Marchenkov_etal_2006}
bottom-up chemical
methods,\cite{Cai_Ruffieux_Jaafar_Bieri_Braun_Blankenburg_Muoth_Seitsonen_Saleh_Feng_etal_2010} 
and carbon nanotubes unzipping.\cite{Jiao_Wang_Diankov_Wang_Dai_2010}
Due to the honeycomb geometry of graphene, GNR can be patterned along two preferential
directions giving rise to armchair shaped edges graphene nanoribbon (aGNR) and zigzag
shaped edges one (zGNR).
The finite size of the GNR gives rise to a large variety of electronic behaviors that could be
relevant in transport.
Considering the case of aGNR, it has been demonstrated that its electronic properties sensibly
vary by changing the width of the ribbon.\cite{Fujita_Wakabayashi_Nakada_Kusakabe_1996,%
PhysRevB.54.17954,PhysRevB.59.8271,Loui06}
In fact, the width and the electron energy gap $\Delta$ are related to each other primarily in inverse
proportion.
In particular, there exist three different classes of $N$-aGNR (aGNR with $N$ dimer lines) for what
concerns the value of the gap $\Delta_{N}$:  $\Delta_{3p+1}>\Delta_{3p}>\Delta_{3p-1}$, with
$p$ integer.\cite{Loui06} 
If we restrict to each of these classes, the energy gap decreases as $p$ increases.
Transport properties of pristine GNR have been studied both for the ideal\cite{PhysRevB.73.195411,%
PhysRevLett.96.246802,PhysRevB.73.235411} and defective\cite{PhysRevB.81.245402,%
Chen_Song_Zhou_Wang_Zhou_2011,Rosales_Orellana_Barticevic_Pacheco_2007} case.
Electronic transport has also been studied, by a tight-binding approach, for junctions connecting
zGNR of different widths\cite{PhysRevB.64.125428,PhysRevLett.84.3390,PhysRevB.74.195417}
revealing the crucial role played by corner edge structures.\cite{Yamamoto_Wakabayashi_2009}
A challenging technological issue would be to exploit the high mobility properties of graphene and the 
finite-size characteristics of GNR, by creating heterostructures which exploit both these advantages. 
These systems may be used as interconnections to transmit signals in future pure-C-based electronic
devices.
In fact, the effects of quantum confinement in graphene nanoconstrictions have been studied showing
their analogy with optics phenomena\cite{PhysRevLett.102.136803} as well as their exploitation for
valley filter applications,\cite{Rycerz_Tworzydlo_Beenakker_2006} and recently a quantized ballistic
conductance has been measured in such structures.\cite{Tombros_2011}
Even more complex junctions have been experimentally built, with carbon nanotubes as
interconnections between graphene bilayers.\cite{Qi_Huang_Feng_Shi_Li_2011}
However, due to their flat structure, GNRs seem easier to pattern than carbon nanotubes.
Owing to the large zoology of hybrid graphene interconnections, it is of great importance to perform
a systematic investigation of their transport properties at the simplest level of configurational
complexity, in order to clarify how the basic geometrical parameters affect the conductance.
Most of the previous works investigating GNR junctions and graphene nanoconstrictions rely on
standard tight-binding calculations. 
It has been shown by Louie and co-workers\cite{Loui06} that the predictions of simple tight-binding
models on GNR may lead to incorrect band structures and energy gaps, since the bonding
characteristics between atoms substantially change at the edges. 
An ab-initio study of GNR junctions avoids such these difficulties, providing an accurate description 
of their transport properties\cite{Roche-nn100028q}, which are directly related to the underlying 
electronic band structure.

In this paper, we present a first-principles study, by means of the non-equilibrium Green's
functions (NEGF) technique, of the electronic and transport properties of systems consisting
of two semi-infinite graphene layers interconnected by an hydrogen-passivated armchair
graphene nanoribbon.
The transport properties of such junctions are predicted to strongly depend on the GNR
geometry, while are quite robust to changes of the graphene's edges geometry.
These structures combine the high mobility of graphene electrodes\cite{PhysRevLett.100.016602,%
Bolotin_Sikes_Jiang_Klima_Fudenberg_Hone_Kim_Stormer_2008,%
Katsnelson_Novoselov_Geim_2006} with the intrinsic semiconducting behavior of GNR.
We show that these semiconducting hybrid graphene-GNR junctions have a significant gap
in the transmission spectrum, which may be exploited to build logic devices which require
a large on-off ratio in the current.\cite{Qi_Huang_Feng_Shi_Li_2011}
In the following discussion, we will show the properties of aGNR junctions considering one
representative for each class according to their width: we will take into account 3-aGNR,
5-aGNR, and 7-aGNR.
In the case of 3-aGNR, we will consider ribbons of three different lengths, namely those
consisting of four (4L), six (6L), and eight (8L) zigzag lines along their axis.
The 8L 3-aGNR will be considered as an illustrative model to study the effects of the
application of a bias.
We will discuss the effect of the rotation of a phenyl ring, and show how different edges
configurations affect the transport properties.

\section{Model and Methods}
\begin{figure}
\includegraphics[width=0.48\textwidth]{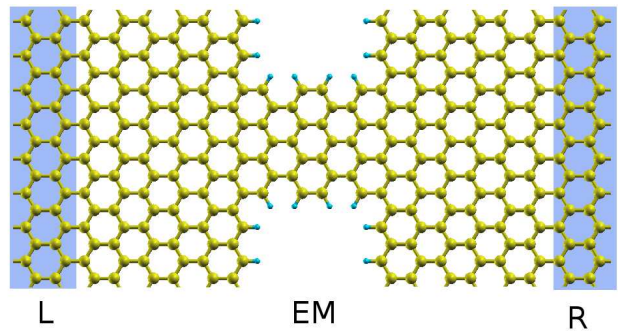}
\caption{\label{fig:setup}
System setup for 4L 7-aGNR.
The left and right electrodes and the extended-molecule region are highlighted.}
\end{figure}
The system setup for a 4L 7-aGNR junction is shown in \ref{fig:setup} as a representative case.
This open system is constituted by three parts: the left (L) and right (R) semi-infinite graphene
leads, and an extended-molecule (EM) region.
The junction is constructed so that the periodic replicas of the aGNR along the direction parallel
to the electrode edge are separated by 7.43~\AA.
This corresponds to three unit cells of graphene and we verified that the interactions among the
replica are negligible.
The width of the leads part included inside the EM region is chosen after converging the
transmission function: six carbon zigzag lines per side are sufficient to reach stable results.

The electronic structure calculations are carried out using the first-principles self-consistent
method implemented in SIESTA package.\cite{Sole02,Arta08}
The exchange-correlation energy and electron$-$ion interaction are described by the
Perdew-Burke-Ernzerhof (PBE)\cite{Perd96} generalized gradient approximation (GGA)
and norm-conserving pseudopotentials\cite{Troullier91} in the fully nonlocal form, respectively.
A double-$\zeta$ polarized basis set of numerical atomic orbitals is used and the energy
cutoff for real-space mesh is set to 200~Ry.\cite{Sole02}
Preliminary tests indicated that the relaxation of the carbon atoms in the leads did not affect
the transport properties of the systems under study, so in most cases we considered
non-relaxed geometries. 
The edges of semi-infinite leads are saturated with one relaxed hydrogen per carbon atom.
We also verified that the relaxation of the aGNR does not affect significantly the electronic
and transport properties of the system. 
For the calculation of the transmission coefficients, 60~$k$-points along the transverse
direction in the 2D first Brillouin zone are used.
Periodic images of the graphene layer are separated by 15~\AA\ along the normal direction.
The electronic transport is studied with the TranSIESTA code,\cite{Bran02} which combines
the NEGF technique with density functional theory.
The transmission function of the system can be obtained by the following equation:
\begin{equation}
T(E,V)=\mathrm{Tr}[\Gamma_{\rm{L}}(E,V)G(E,V)\Gamma_{\rm{R}}(E,V)G^{\dag}(E,V)],
\end{equation}
where the spectral density $\Gamma_{\rm{L}\rm{(R)}}$ describes the coupling between the
L (R) electrode and the EM region and it is given by the imaginary part of the electrode
self-energy: $\Gamma_{\rm L(R)}(E,V)=i(\Sigma_{\rm L(R)}-\Sigma_{\rm L(R)}^{*})/2$.
The self-energy describes the hopping across the surface separating one lead and the EM
region, and establishes the appropriate boundary conditions for the Green's function
calculation.
$G$ is the retarded Green's function of the EM region, formally given by:
\begin{equation}
G(E,V)=\left[ES-H(V)-\Sigma_{\rm{L}}(E,V)-\Sigma_{\rm{R}}(E,V)\right]^{-1},
\end{equation}
where $S$ is the overlap matrix and $H$ is the Hamiltonian of the system when a bias
voltage $V$ is applied.
The current is simply given by the following integral:
\begin{equation}
I(V)=\frac{2e^{2}}{h}\int_{-\infty}^{\infty}\mathrm{d}E\ T(E,V) \left[ f(E-\mu_{\rm{L}})-
f(E-\mu_{\rm{R}})\right],
\end{equation}
with $f$ being the occupation Fermi function, $\mu_{\rm{L}\rm{(R)}}$ the chemical
potential of the L (R) electrode, and $V=\mu_{\rm{L}}-\mu_{\rm{R}}$.

\section{Results and discussion}
\begin{figure}
\includegraphics[width=0.48\textwidth]{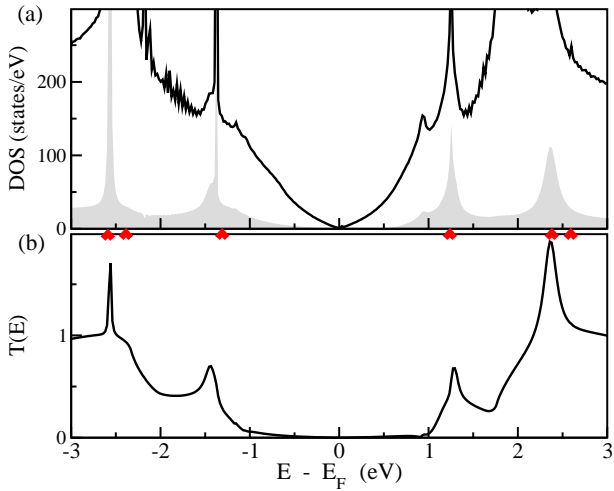}
\caption{\label{fig:dos-te}
Electronic and transport properties of 4L 3-aGNR.
Upper panel (a): DOS (solid line) of the junction and PDOS (shaded grey area) on the ribbon
region.
Red triangles represent the eigenstates of the isolated (unconnected and hydrogenated)
4L 3-aGNR linker.
Lower panel (b): transmission function.}
\end{figure}
We first investigate in detail the electronic properties and the transmission function of the 4L
3-aGNR junction.
For the other systems, similar considerations can be done.
In \ref{fig:dos-te} we show the density of states (DOS) in the EM region and the projected
density of states (PDOS) on the ribbon region for the 4L 3-aGNR, in comparison with the
transmission function.
Around the Fermi energy ($E_{\mathrm F}$), the DOS resembles that of graphene, in fact
it goes to zero almost linearly with a deviation due to the presence of vacuum portions in
the molecular bridge region.
The signature of the molecular energy levels is clear both in the DOS and PDOS, and the
transmission function has higher intensity in correspondence to those peaks.
We also consider the isolated molecule obtained by cutting the bonds between the 4L
3-aGNR and the graphene leads, and saturating them with hydrogen atoms.
The red triangles represent the eigenvalues of the isolated molecule, which correlate well
with the position of the resonances in the DOS, PDOS, and transmission.
The (P)DOS for the other studied systems have a similar behavior, and are not reported here.
\begin{figure}
\includegraphics[width=0.48\textwidth]{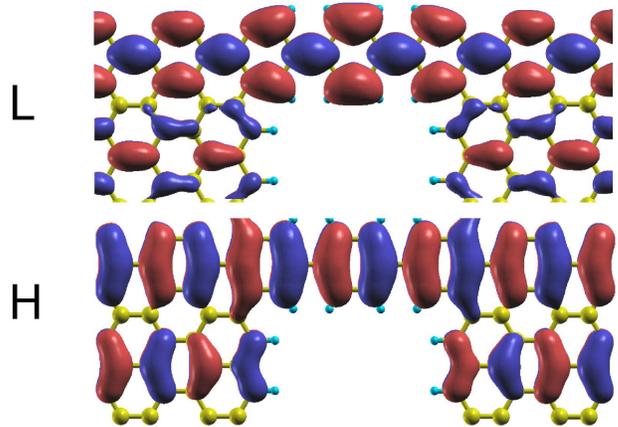}
\caption{\label{fig:autofunz}
4L 3-aGNR junction: isosurface plot of the wavefunctions at energies corresponding to the
LUMO and HOMO resonant transmission peaks, calculated at $k_\parallel=0$.}
\end{figure}
\begin{figure}
\includegraphics[width=0.4\textwidth,angle=-90]{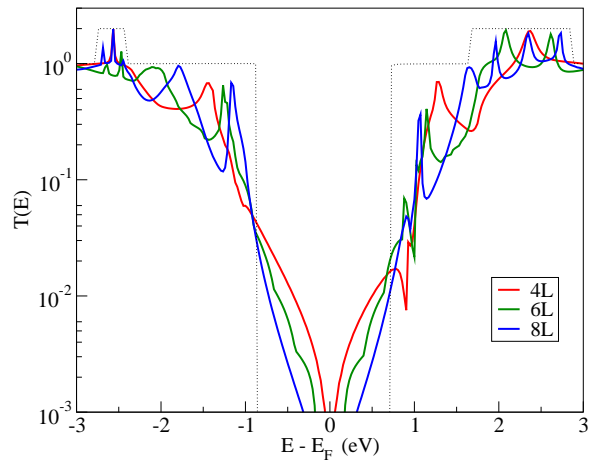}
\caption{\label{fig:3aGNR-te}
Transmission function of 3-aGNR junctions for three different lengths: 4L (red), 6L (green),
and 8L (blue).
The dotted line is the transmission function of the pristine infinite 3-aGNR.}
\end{figure}

\begin{figure}
\includegraphics[angle=90,width=0.24\textwidth]{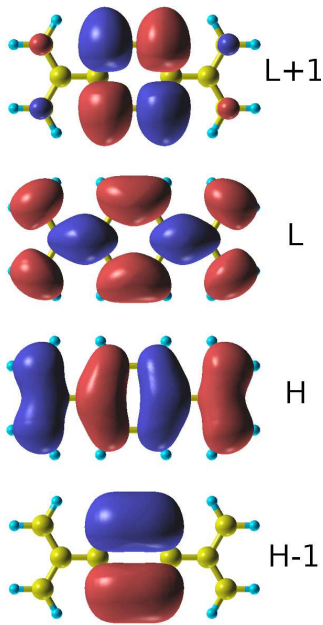}
\caption{\label{fig:wfiso}
Isosurface plots of the selected molecular orbitals of the isolated 3-aGNR linker: HOMO (H),
HOMO-1 (H-1), LUMO (L) and LUMO+1 (L+1).}
\end{figure}

Moving away from $E_{\mathrm F}$, the transmission function increses up to significant
values in correspondence to the two peaks at $-$1.2 and 1~eV; these peaks are
generated by the hybridization of the molecule's highest occupied molecular orbital (HOMO)
and the lowest unoccupied molecular orbital (LUMO) states with the leads, playing the role
of channels between the two graphene electrodes.
The corresponding wavefunctions of the interacting system (shown in \ref{fig:autofunz})
calculated at $k_\parallel=0$, have the same shape and symmetry of that of the isolated
molecule (see \ref{fig:wfiso}).
These wavefunctions represent good conducting states, as they are delocalized along the
molecule and propagate inside the electrodes with the same symmetry.
Some other states farther from $E_{\mathrm F}$ contribute to the transmission.
All these states are generated from the $\pi$ orbitals of the C atoms and they are delocalized
along the junction.

We now discuss the dependence with the length of the ribbon forming the junction.
The transmission functions for the 4L, 6L, and 8L 3-aGNR junctions, reported in
\ref{fig:3aGNR-te}, show some general trends.
The peaks become denser when the length of the junction increases, because they are
related to the discrete structure of the electronic states of the nanostructured ribbon.
Thus, the effective gap of the junction decreases as the length increases.
As expected, the transmission function of the pristine infinite 3-aGNR acts like an envelope
curve for the other curves; it has an energy gap of $\simeq$1.5~eV.\cite{Loui06}
This aspect confirms that the contacts between the graphene sheets and the nanoribbon do
not represent significant barriers to the transport of electrons, as one can naively expect due
to the chemical identity of the different subsystems.
Within the energy gap of the nanoribbon, the transmission function is very low and decreases
rapidly as the length of the ribbon increases.
This behavior is clearly due to the tunneling mechanism dominating the transmission at those
energies.
The states of the graphene lead decay exponentially along the ribbon because there are no
states in the junction supporting the conductance.
This metal-semiconductor-metal device presents an effective gap for transport, since the
conductance inside the gap is several orders of magnitude lower than outside.
\begin{figure}
\includegraphics[width=0.4\textwidth,angle=-90]{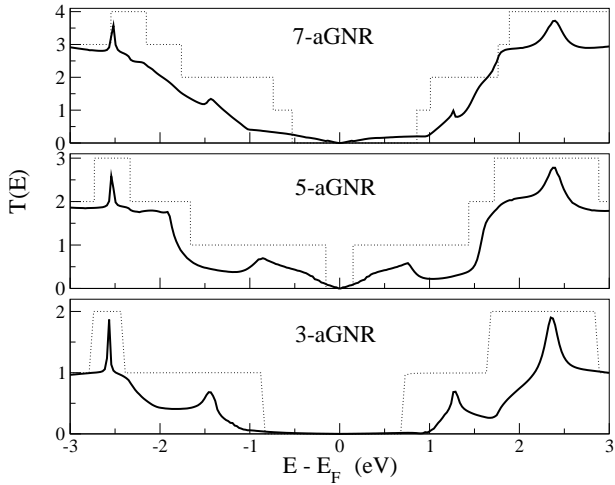}
\caption{\label{fig:width}Transmission function of 4L $N$-aGNR junctions for $N=3$, 5, and 7.
Dotted line: transmission function of the infinite pristine $N$-aGNR.}
\end{figure}

We now discuss the effects of the aGNR width in the transmission function, considering
junctions made with 4L 3-aGNR, 4L 5-aGNR, and 4L 7-aGNR.
We consider a representative aGNR for each of the three classes,
so that we can capture the main differences among them.
As illustrated in \ref{fig:width}, the aGNR junctions have different gaps in the transmission
spectrum consistently with the previously reported results.\cite{Loui06}
In miniaturized graphene-based electronic devices these structures may act both as linkers and
as active components, and it would be useful to finely tune their width in order to have different
conductive behaviors.
Within each class, we expect the energy gap to decrease as the width of the junction increases.
We notice that the intensity of the transmission function is higher for wider ribbons, as more
electronic channels are open.
Moreover, as the width of the ribbon increases the spike features of the transmission function
become less marked, since the system approaches the limit of graphene.
In this case, the transmission functions of the pristine aGNR is a good reference only for those
junctions where the aspect ratio between the length and the width is reasonably high.
This because the contribution to the conductance of the infinite aGNR comes from
$k_\parallel=0$ only.  
For finite nanoribbons this restriction is relaxed and the passage of electrons with a finite
$k_\parallel$ is allowed, with $k_\parallel$ within an interval that increases  as the width to length
ratio increases.
This is clearly reflected in the transmission function, which departs from an exponential decay
within the energy gap for short and wide ribbons.
This is more evident for the 4L 7-aGNR in \ref{fig:width}.
For this ribbon the conductance is linear in the gap region,  reflecting the underlying electronic
structure of the graphene electrodes, and departing from a strictly tunneling behavior dominated
by the energy dependence of the effective tunneling barrier.

We now discuss the effects of the application of a bias to the junction.
We consider the junction with the 8L 3-aGNR as a representative case; this is the longest
junction considered here, thus being that in which the edges of the two graphene leads
are less interacting.
\begin{figure}
\includegraphics[width=0.48\textwidth]{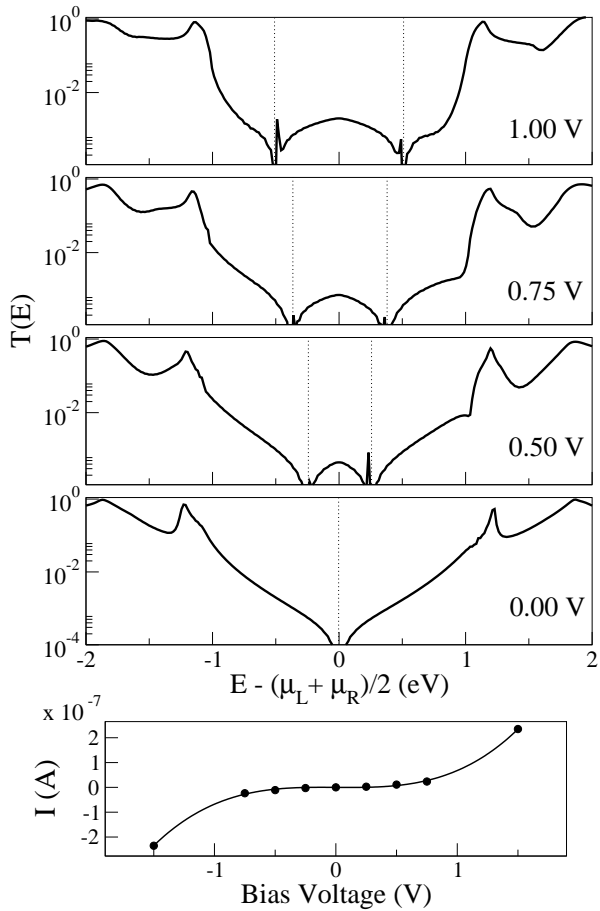}
\caption{\label{fig:bias} Lower panel: current-voltage characteristics of 8L 3-aGNR junction.
Upper panels: transmission function of 8L 3-aGNR junction for four selected values of the bias
voltage, as indicated in labels. The bias window is delimited by two vertical lines.}
\end{figure}
In \ref{fig:bias} the transmission functions for different biases up to 1.0~V are shown.
For any given bias voltage, the chemical potentials of the two leads are well recognizable,
as they correspond to the left and right Dirac points where the DOS tends to zero.
Within the bias window, the intensity of the transmission function increases with the bias.
This is expected since it is determined by the number of electron tunneling from the leads, and
the DOS of graphene shows a linear energy dependence in that region.
Thus, the shape of the transmission function within the gap region can be interpreted in terms
of the product of the DOS of the left an right graphene leads and a modulation function
determined by the size of the gap of the nanoribbons.
We notice two small peaks near the Dirac points of the two electrodes, which we assign to the
edge states appearing at the two zigzag
electrodes.\cite{Wassmann_Seitsonen_Saitta_Lazzeri_Mauri_2008}
In fact, in presence of zigzag edges there is a band near $E_{\mathrm F}$ which corresponds
to a state localized at the edge.
At small applied bias, the probability for an electron lying in an edge state to be transferred to
the opposite lead is very small since the DOS at that energy is negligible.
However, at finite bias, the situation is different and we can find distinct features in the
transmission curves related to the presence of edge states.
We can also see that the resonances originated by the HOMO and LUMO of the aGNR slightly
shift as the bias is applied.
As a result, the energy gap reduces by approximatively 10\% when a bias of 1.0~V is applied.
We verified that the current remains very low (less that 0.1~$\mu$A) for biases up to 1.0~V.
For larger values, the intensity sharply increases when the two main peaks enter the bias
window, and for a bias of 3.0~V we calculated a current of 3.3~$\mu$A.

\begin{figure}
\includegraphics[width=0.48\textwidth]{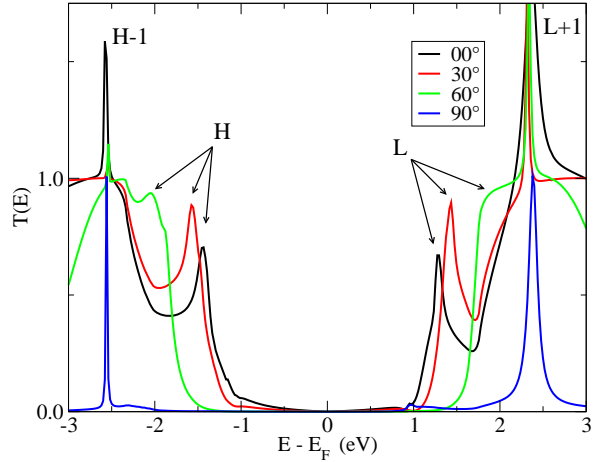}
\caption{\label{fig:transm-tors}
Transmission function at zero bias for the 4L 3-aGNR junction, under rigid torsion of the central
phenyl ring.
The curves shown here refer to rotations by $0^{\circ}, 30^{\circ}, 60^{\circ}$, and $90^{\circ}$.
The main resonances are highlighted and the labels correspond to the 3-aGNR molecular
states shown in \ref{fig:wfiso}. }
\end{figure}

When considering junctions made of 3-aGNR, the system is actually a chain of phenyl rings.
In this case, it has been shown that the stable configuration consists of neighbouring rings lying
on different planes.\cite{Lortscher_Elbing_Tschudy_VonHanisch_Weber_Mayor_Riel_2008,Venkataraman_Klare_Nuckolls_Hybertsen_Steigerwald_2006}
We can thus take into account one more degree of freedom, that is the relative torsion angle
between the phenyl rings of the linkers.
We show the effect of the ring rotation by taking the 4L 3-aGNR as test system.
According to our calculations, the most stable configuration corresponds to a rotation of
$\simeq 45^{\circ}$.\cite{Vergniory_Granadino-Roldan_Garcia-Lekue_Wang_2010}
In fact, the central ring tends to rotate with respect to its planar configuration due to steric
repulsion between its hydrogen atoms and those of the neighboring rings.
By looking at the transmission function in \ref{fig:transm-tors} we can note a general trend.
As the ring rotates from $0^{\circ}$ to $90^{\circ}$, the peak H (HOMO) shifts towards lower
energies, while the peak L (LUMO) shifts towards higher energies.
As a result, the energy gap of the system increases.
In fact, the wavefunctions related to these peaks (see \ref{fig:wfiso}) are extended along
all the molecular backbone and they feel a distortion as the central ring is rotated, thus their
energies are expected to change.
Both HOMO and LUMO are even respect to the mirror plane that bisects the 3-aGNR molecule.
As consequence of this symmetry, when the rotation is by $90^{\circ}$ these two peaks
disappear: the linear combinations of the ${p_z}$ orbitals of the central carbon atoms are
orthogonal to those of the remaining carbon atoms, resulting in the closure of the
corresponding conduction channels.
The situation is different for HOMO$-$1 and LUMO+1, whose wavefunctions are mainly
localized on the central phenyl ring.
In fact, they are at the same energy position regardless of the torsion angle.
These considerations may be extended to the case of longer aGNR.

\begin{figure}
\includegraphics[width=0.48\textwidth]{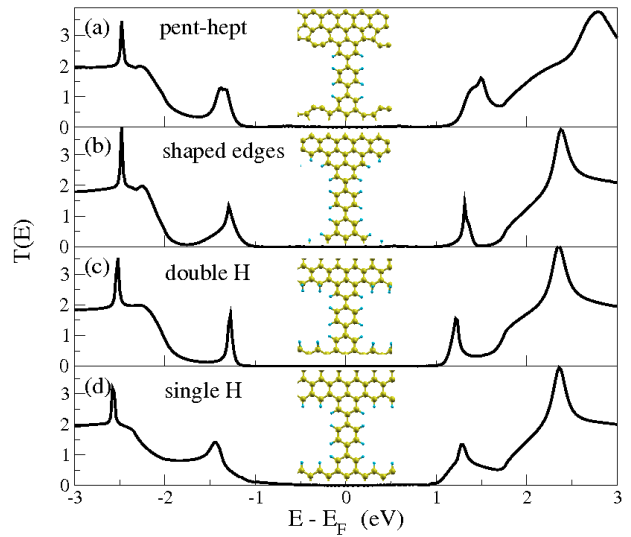}
\caption{\label{fig:sat}
Zero bias transmission function of 8L 3-aGNR junctions with four different saturations of the
graphene edges.
The inset show the linking geometries.}
\end{figure}

Concerning the passivation of the graphene edges, all the results presented so far refer to
what we can call ``standard saturation''.
That is, each edge carbon atom is saturated by a single hydrogen (panel $a$ in \ref{fig:sat}).
In a real system subject to different hydrogen partial pressures, one may guess that the edge
saturation can change.\cite{Wassmann_Seitsonen_Saitta_Lazzeri_Mauri_2008}
In order to validate our results, we show the effect of four different edge saturations on the
transmission function.
The first case we consider is the formation of pentagon-heptagon reconstruction of the
graphene edges, where no H atoms are present on the edge (panel $d$).
Then, we considered the case in which each carbon edge atom is saturated by two
hydrogens (panel $b$).
One may also guess a more complicated connection between the aGNR and the leads, and
we modeled it as shown in the inset of panel $c$.
\ref{fig:sat} shows the transmission function for these four cases.
Our results are very robust with respect to these changes of the geometry and saturation of
the graphene lead edges, as we see only minor differences among these cases, mainly a
sharpening of the peaks derived from the HOMO and LUMO resonances.
This is evident in particular for the double-H and shaped edges, in which the orbitals of the
aGNR are less hybridized with the leads.
The reason is that in both cases less charge is present on the graphene layer edge.

\section{Conclusions}
Very simple prototypes of carbon-based molecular junctions have been investigated in
order to characterize their transport properties.
Carbon nanoribbons with armchair shaped edges have been considered as linkers between
between two semiinfinite graphene electrodes.
These systems present a typical metal-semiconductor-metal behavior due to the electronic
gap of the ribbons which depends on the their width.
The electronic properties of the isolated subsystems, i.e. graphene electrodes and the
nanoribbon, together with the interaction in contact regions determine the transport properties
of the junction.
For what concerns the coupling between the subsystems, this is very efficient and the contacts
do not create appreciable barrier for transport due to the same chemical species constituting
the subsystems.
Hence the transport properties are mainly determined by the shape (width and length) of the
finite ribbon included in between the graphene leads.
Larger molecules furnish more channels for convoying the electrons then the smaller ones and
consequently the conductance is generally higher.
In the energy gap region the transport occurs via electron tunneling between the electrodes
and the efficiency decays simply with the linker length.

Some additional representative configurations has been considered explicitly: thinest ribbon
a torqued geometries has been studied to evaluate the effects of the misalignment of the
$\pi$ orbitals on different phenyl rings.
The changes in transmission through HOMO and LUMO orbitals are noteworthy, while for energy
levels farther from $E_{\textrm F}$ they are negligible.
Finally, graphene edges with different saturations and reconstructions has been considered and
they do not influence significantly the calculated transport properties.
In spite of the simple model adopted in this paper, the reported results on the electronic transport
provide some parameters to control and engineer future carbon-based electronic devices.


\section{Aknowledgments}
C.M. thanks CARIPLO Foundation for its support within the
PCAM European Doctoral Programme.
DSP acknowledges support fromt from Basque Departamento de Educaci\'on, 
UPV/EHU (Grant No. IT-366-07), the Spanish Ministerio de 
Ciencia e Innovaci\'on (Grant No. FIS2010-19609-C02-02), and 
the ETORTEK research program funded by the Basque Departamento 
de Industria and the Diputaci\'on Foral de Gipuzkoa.

\bibliography{gnr-graphene-prb}
\end{document}